\documentclass{aa}
\usepackage{times,graphics,psfig}
\def\hera{Hercules A }
\begin{document}
\thesaurus{03(11.01.2 - 11.09.1:Hercules A - 13.25.3 - 13.18.1)}
\title{Spatial and spectral X-ray properties of the powerful radio galaxy
Hercules A: Environment and jet/ICM interaction}
\author{J. Siebert\inst{1,2}, N. Kawai\inst{2}
\and W. Brinkmann\inst{1}}
\mail{J. Siebert, jos@mpe.mpg.de}
\institute{Max--Planck--Institut f\"ur extraterrestrische Physik,
 Giessenbachstrasse, D--85740 Garching, Germany
\and
 Institute of Physical and Chemical Research (RIKEN), 2-1 Hirosawa, Wako,
 Saitama 351-01, Japan}
\titlerunning{Hercules A}
\authorrunning{Siebert et al.}
\date{Received 17 May 1999; accepted}
\maketitle
\begin{abstract}
We investigate the X-ray properties of the powerful radio galaxy
Hercules A (3C 348) using ROSAT HRI, PSPC and ASCA observations.
The ASCA data are well fit by a thermal plasma model with 
a temperature of about 4.3 keV and abundances 0.4 solar. The HRI 
observation clearly reveals extended and elongated X-ray emission. 
For radii greater than 10 arcsec, the surface brightness profile 
perpendicular to the elongation is well fit by an isothermal 
$\beta$--model with $\beta = 0.63$ and a core radius of $\approx 120$ 
kpc. We derive a total mass of the putative cluster of $8.4\times 10^{13}$
M$_{\sun}$ and a gas mass fraction of about 18 per cent. The inner 
part of the surface brightness profile clealry reveals the presence 
of a point source, which contributes $\approx$8 per cent to the total 
flux. The 0.1-2.4 keV luminosity of the point-like and the extended emission 
is $3.4\times 10^{43}$ and $4.3\times 10^{44}$ erg s$^{-1}$, respectively. 
After subtracting the cluster X-ray emission from the HRI image, 
residual structures are visible, which partly coincide with the radio 
jet and lobes. This indicates an interaction of the radio jet with the
intracluster medium.  
\keywords{Galaxies: active; Galaxies: individual -- 
Hercules A; X--rays: general -- Radio continuum: galaxies.}

\end{abstract}

\section{Introduction}

The giant elliptical radio galaxy \hera (3C 348, 1648+05, z = 0.154) is 
one of the most prominent AGN in the sky with a total radio flux density of 
351 Jy at 178 MHz, which makes it the fourth brightest extragalactic radio 
source. It was one of the first optically identified extragalactic radio sources 
(Bolton 1948) and has been studied extensively in the radio and optical since then. 

A high resolution radio map of \hera at 5GHz (Dreher \& Feigelson 1984) 
shows giant radio lobes and prominent jets with spectacular (and still unexplained)
ring-like structures. The classification of \hera in terms of Fanaroff \& Riley (FR) 
type is ambiguous. Despite its extremely high radio power (log P$_{\rm tot,5GHz}$ = 
27.19 W/Hz), which is typical for FR II radio galaxies, the radio morphology argues 
for a FR I classification (no hot spots visible in the radio lobes, prominent 
two-sided jets).
 
Broadband optical imaging revealed two concentrations of light. Thus, 
\hera was originally thought to be a cD galaxy with a foreground star 
superimposed on it (Greenstein 1962), although already Minkowski (1957) 
classified it as a double galaxy. Recent photometric work by Sadun \& Hayes 
(1993) supports the idea of a double nucleus with a separation of 
$\sim 4^{\prime\prime}$, thus indicating merging activity. 

The environment of \hera has been investigated by several authors (Greenstein 
1962, Yates et al. 1989, Allington-Smith et al. 1993), but 
whether or not \hera is associated with a cluster is not yet unambiguously decided. 
Whereas Yates et al. (1989) claim no particularly rich environment around Hercules A, 
Allington-Smith et al. (1993) find an excess number of galaxies within 500 kpc of 
Hercules A. It has to be noted, however, that all studies were based on statistical 
methods, i.e. number counts, and there is no spectroscopic confirmation of a cluster
of galaxies up to now.

In X-rays \hera was observed for the first time with the 
{\it Einstein} observatory. Dreher \& Feigelson (1984) note extended emission on 
scales of $\ge$ Mpc and a luminosity of 3.4$\times$ 10$^{\rm 44}$ erg s$^{\rm -1}$ 
in the 0.2--4 keV energy band. However, a detailed analysis of the X-ray data has never 
been published. 

In this paper we present a thorough analysis of the X-ray properties of \hera using
archival ROSAT data and new ASCA observations. We will show that the X-ray emission
is extended and predominently thermal and that there is interaction between the radio 
jet and the X-ray emitting gas. The paper is structured as follows: In Sect.2 we analyse 
the ASCA and ROSAT PSPC data and determine the spectral properties of Hercules A. 
In Sect.3 we investigate the spatial structure of the X-ray emission using a ROSAT HRI 
observation and we compare it to the VLA radio map. A discussion of the results is 
given in Sect.4 and our conclusions are presented in Sect.5. Spatial scales and 
luminosities are calculated assuming $H_{\rm o} = 50$ km s$^{-1}$ Mpc$^{-1}$, $q_{\rm o} = 
0.5$ and $\Lambda = 0$ throughout this paper. At the distance of Hercules A one arcmin 
corresponds to $\approx 209$ kpc.  

\section{Spectral analysis}

\subsection{The ASCA and ROSAT PSPC observations}

The ASCA observation was performed on August 12 and August 13 1998 in 1-CCD faint mode.   
The data were analysed using {\footnotesize FTOOLS 4.1}. The recommended standard screening 
criteria were applied. 

In addition, data taken within 60 seconds after
passage of the day-night-terminator and the South Atlantic Anomaly (SAA) were 
not considered in the analysis. Periods of high background
were manually excluded from the data by checking the light curve of the
observation. The resulting effective exposures were 35.1 and 34.4 ksec for GIS2 and GIS3
and 33.4 and 32.9 ksec for SIS0 and SIS1, respectively.

Source counts were extracted from a circular region centered on the target 
with a radius of $6\arcmin$ for the GIS and $4\arcmin$ for the SIS.
We used the local background determined from the observation in the
analysis for both detectors. In particular, the GIS background was estimated 
from a source free region at the same off-axis angle as the source and with 
the same size as the source extraction region. In total, $\sim$4000 source 
photons were detected in each of the SIS detectors. 

All spectra were rebinned to have at least 20 photons in each energy channel. 
This allows the use of the $\chi^2$ technique to obtain the best fit values 
for the model spectra. We used the latest GIS redistribution matrices 
available (V4\_0) from the calibration database and created the SIS response 
matrices for our observation using {\footnotesize SISRMG}, which applies the
latest charge transfer inefficiency (CTI) table ({\tt sisph2pi\_110397.fits}).
The ancillary response files for all four detectors were generated using the 
{\footnotesize ASCAARF} program. 

Spectra were fitted in the energy range 0.8 to 9 keV for both GIS. For SIS0 
and SIS1 we used 0.8 to 8 keV. The upper energy boundaries are given by the 
maximum energy at which the source was detected in each instrument. The lower 
energy boundaries result from the calibration uncertainties of the detectors. 
In particular, SIS1 recently shows systematic residuals below 0.6 keV (Dotani 
et al. \cite{dotani}). Also, since the CCD temperature of both SIS detectors 
was rather high during the observation, we decided to ignore all data below 0.8 keV.
An increasing RDD (Residual Dark Distribution), which is caused by radiation 
damage, has been reported for both CCDs. This effect cannot be corrected for 
with currently available software. However, RDD degradation should be negligible 
for 1-CCD observations and affects mostly the lowest energy channels, which are
excluded from our analysis anyway (Dotani et al. \cite{dotani}).

The ROSAT PSPC data (ROR 701611) were obtained from the public data archive at MPE. 
\hera was observed with the ROSAT PSPC between August 19 and August 31 1993 with 
an effective exposure of $8058$ sec. The data were prepared for spectral analysis 
using standard commands within the EXSAS environment. In short, photons were extracted 
from a circular region with a radius of 5 arcmin 
centered on the peak of the X-ray emission. The background was determined from a source-free 
annulus with inner radius 6 arcmin and outer radius 9 arcmin. Only pulse height channels 
12 to 240 were used for spectral analysis, because of calibration uncertainties in the 
lowest energy channels. The data were rebinned to get a signal-to-noise ratio of at least 
five in each energy bin. Finally, vignetting and dead time correction were applied to the 
binned dataset.

\subsection{Results}

\begin{figure}
\hspace*{1cm}
\psfig{file=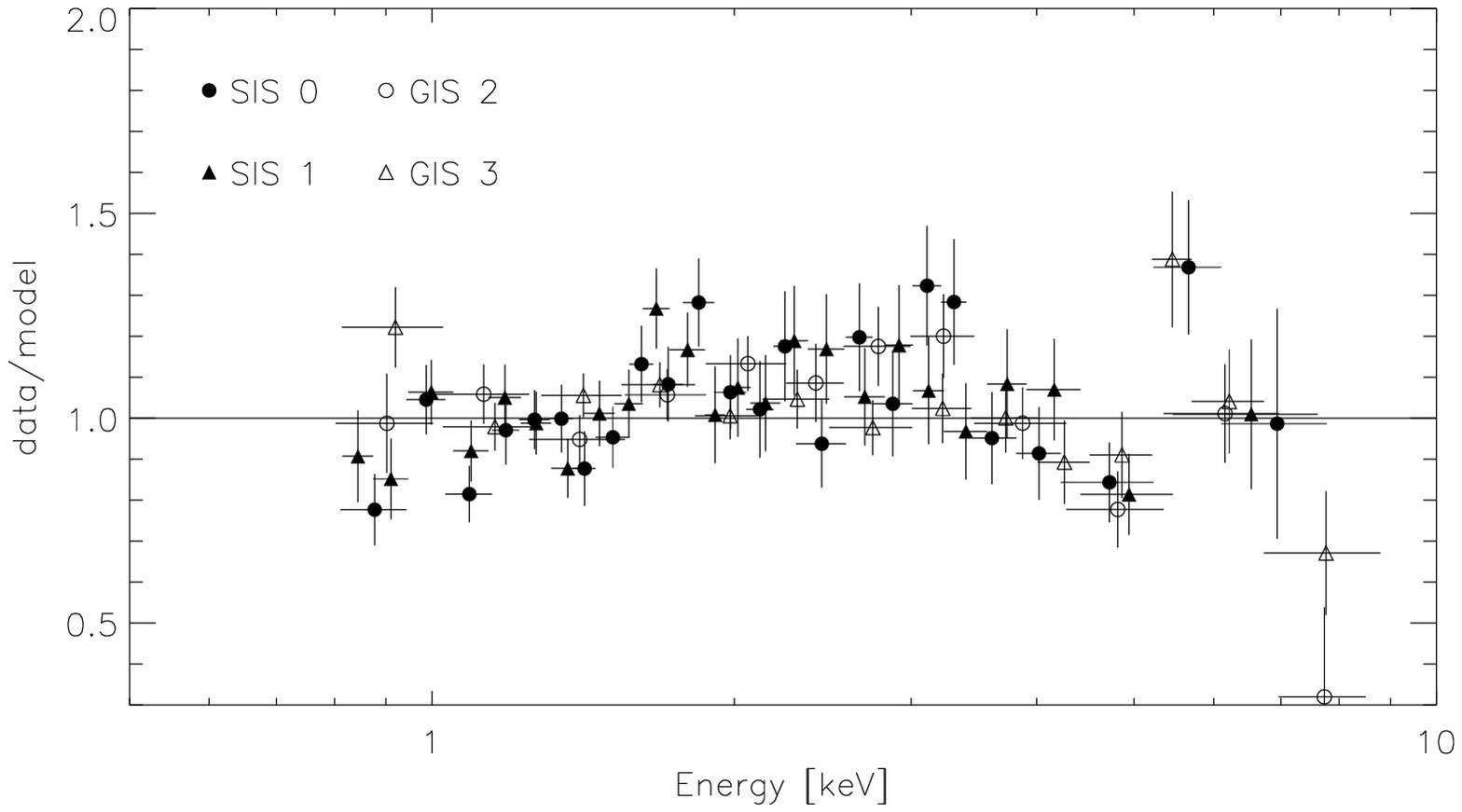,width=7.5cm}
\vskip0.5cm\hspace*{1cm}
\psfig{file=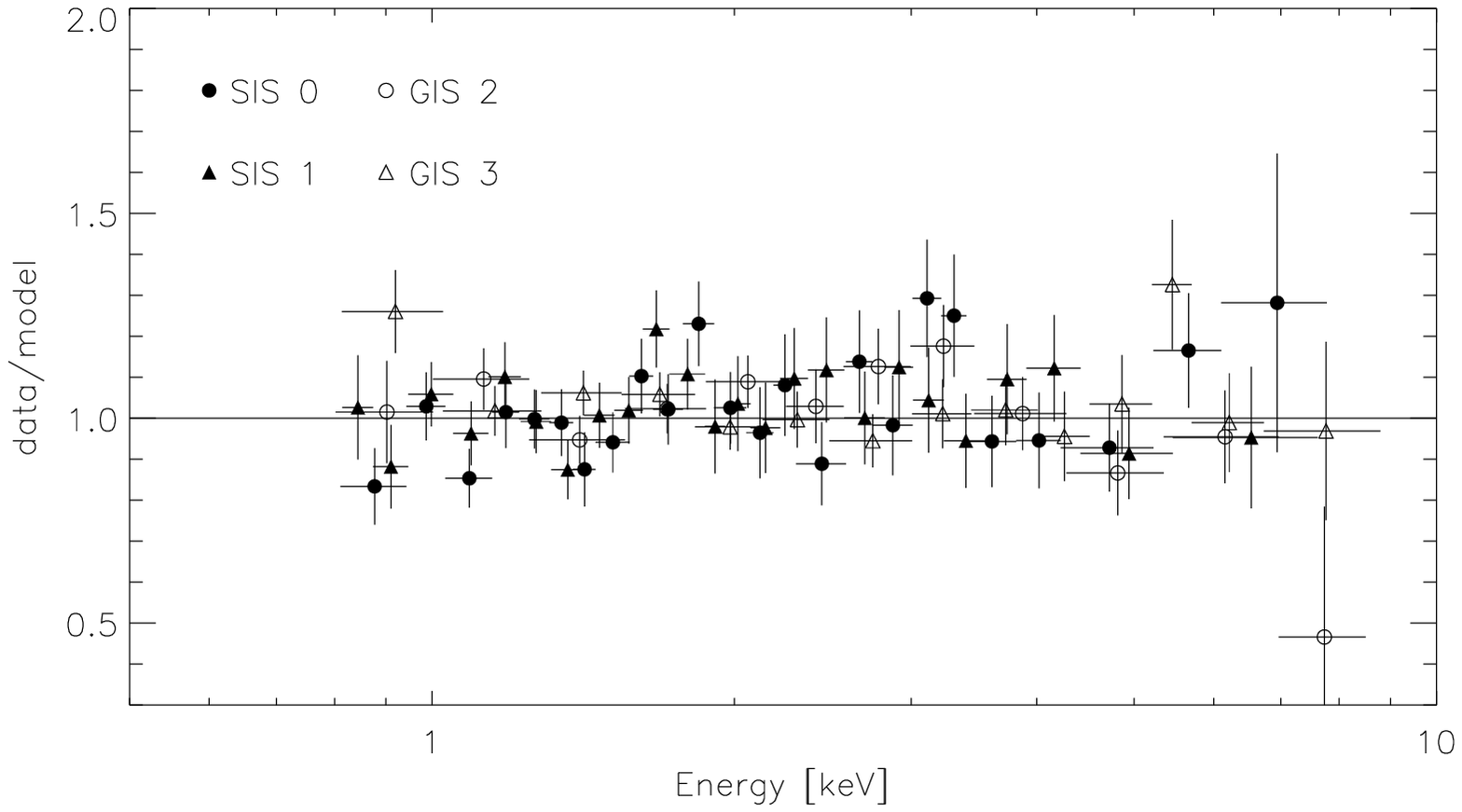,width=7.5cm}
\vskip0.4cm
\caption{Ratio of the ASCA data from all four detectors to a simple
power law model (top panel) and the best-fit thermal model including
a five per cent power law contribution from the AGN (bottom panel). The data
are rebinned to match the energy resolution of the instruments.}
\label{ascaspec}
\end{figure}

We first investigated the data from each detector individually and the results
agreed within the errors. In the following we therefore only cite the best-fit
model parameters of a simultaneous fit to all four instruments.

The ratio of the ASCA data to a simple power law model with Galactic absorption
($N_{\rm H} = 6.40\times 10^{20}$ cm$^{-2}$) is shown in the upper panel of 
Fig.~\ref{ascaspec}. Clearly, there are systematic residuals, in particular above 
3 keV. For completeness, we give the best-fit photon index: 
$\Gamma = 1.91^{+0.04}_{-0.03}$ ($\chi^2 = 645.4$ (586 d.o.f.)).

Next we tried a thermal emission model, namely the {\small \sc Mekal} model within
XSPEC. The fit improves dramatically ($\Delta\chi^2 = -61.7$ for one additional
parameter). The best-fit parameters are as follows: $kT = 4.62^{+0.27}_{-0.25}$ keV,
$Z = 0.38\pm0.10 Z_{\sun}$ and $A = (5.0\pm0.2)\times 10^{-3}$ photons cm$^{-2}$ s$^{-1}$ 
keV$^{-1}$. Since the spatial analysis indicates a small contribution
from the central AGN (see Sect.3), we finally included a power law component and fixed its
normalization to 10 per cent of the thermal emission. The fit does not further
improve ($\Delta\chi^2 = +1.1$ for one additional parameter), but the model is well 
consistent with the data ($\chi^2_{\rm red} = 1.00$). The best-fit parameters of
this model are $kT = 4.25^{+1.00}_{-0.66}$ keV, $Z = 0.44\pm 0.13 Z_{\sun}$, 
$\Gamma = 1.71^{+0.74}_{-0.30}$ and $A_{\rm th} = 4.05^{+0.14}_{-0.17}\times 10^{-3}$
photons cm$^{-2}$ s$^{-1}$ keV$^{-1}$. The corresponding data to model ratio is shown 
in the lower panel of Fig.~\ref{ascaspec}. 

If the above model is applied to the PSPC data with all parameters fixed to their
best-fit values, we only get a marginally acceptable description of the PSPC spectrum
($\chi^2_{\rm red} = 1.37$ (37 d.o.f.)). However, the fit significantly improves for 
lower temperatures ($kT = 2.44^{+1.33}_{-0.60}$ keV; all other parameters fixed to 
the ASCA values). This might indicate that a single temperature thermal plasma is not
a good description for the intracluster medium around Hercules A. In the case of a
more complex temperature structure, the relative contribution of the lower temperature 
plasma in the ROSAT band is higher than in the ASCA band. Therefore the PSPC observation
is more sensitive to the lower temperature component, whereas the ASCA measurement
is dominated by the higher temperature component. However, the fact that the best-fit
temperature determined from the PSPC data is close to the upper end of the PSPC energy
range might also indicate that instrumental effects and maybe even residual ROSAT-ASCA
cross-calibration uncertainties contribute to the difference in the obtained
plasma temperatures from the two instruments.               

We further investigated if there is a radial temperature gradient, but found none.
Fitting the PSPC data from the inner 45 arcsec and from an annulus with inner radius
45 arcsec and outer radius 3 arcmin seperately, we do not find significantly different
temperatures ($kT_{\rm in} = 2.17^{+0.83}_{-0.41}$ keV compared to $kT_{\rm out} = 
2.68^{+1.74}_{-0.72}$ keV). Also the (lower resolution) ASCA data do not indicate 
any temperature gradient.

\section{Spatial analysis}

\subsection{The HRI observation}

\hera was observed with the ROSAT {\it High resolution Imager} (HRI) between August 28
and September 11 1996 ( ROR 702755). The effective exposure of this observation was 
21.52 ksec. A second, much shorter (1.2 ksec) observation is not considered in this
analysis. The total count rate for \hera is $0.077\pm 0.010$ cts/s. 

\begin{figure}
\psfig{file=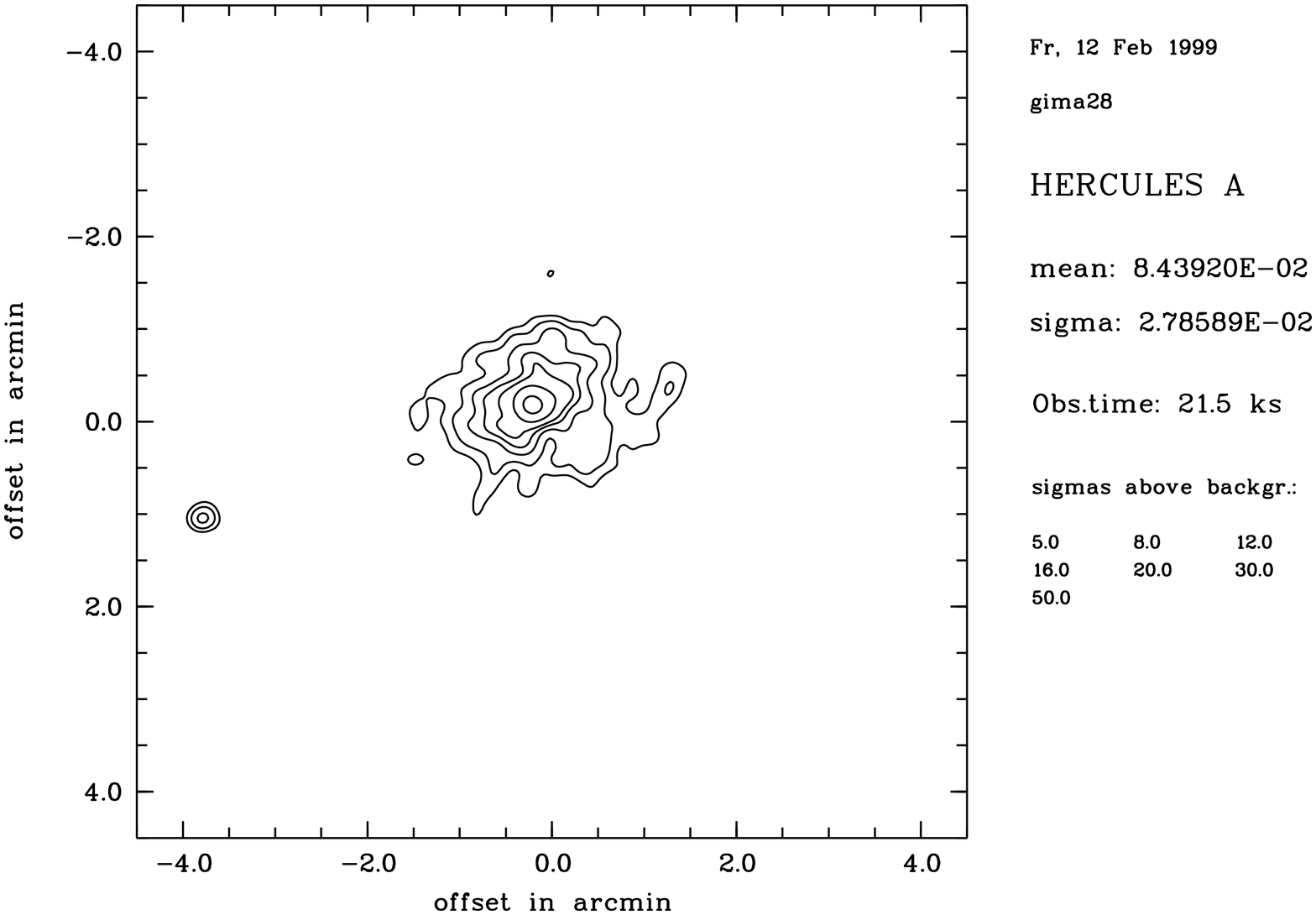,width=8.5cm,bbllx=45pt,bblly=70pt,bburx=552pt,bbury=605pt,clip=,angle=-90}
\caption{X-ray contours of Hercules A. The contours correspond to 5,8,12,16,20,30 and 50 
$\sigma$ above background. The lowest and highest contours denote intensity levels of
$9.4\times 10^{-3}$ and $61.8\times 10^{-3}$ cts s$^{-1}$ arcmin$^{-2}$ respectively.} 
\label{cont}
\end{figure}

Only HRI pulse-height channels 2 to 8 are used in the spatial analysis in order to 
increase signal-to-noise, as channels 9 to 15 mostly contain instrumental background 
(cf. David et al. 1997).
A contour plot of the total X-ray emission is shown in Fig.~\ref{cont}. The photons 
were binned in $2\times2$ arcsec pixels and subsequently smoothed with a Gaussian with 
$\sigma = 6$ arcsec.

The X-ray emission is clearly extended and elongated. As will be shown later, the 
direction of the elongation is close to the radio axis of Hercules A. We are confident
that the peculiar shape of the X-ray emission is indeed intrinsic and not an 
instrumental effect, {\bf because} there is a second X-ray source about 4 arcmin to the 
south-east of Hercules A (R.A. = 16h51m22.3s, Dec = +04d58m23s), 
which is point-like. It is associated with a stellar object of $m_{\rm V} = 12.77$ on 
the digitized POSS-I plate. 

\subsection{Surface brightness profile}

To determine the physical properties of the extended X-ray emission we fitted a 
$\beta$--model (e.g. Cavaliere \& Fusco-Femiano 1976; Jones \& Forman 1984) of the
form
\[
S(r) = S_0 \left(1+ \frac{r^2}{r_c^2}\right)^{-3\beta+1/2}
\]
to the surface brightness profile of the HRI source. However, we did not use an
azimuthally averaged radial profile, but extracted photons from a cone with an opening
angle of 90 degrees oriented perpendicular to the elongation of the X-ray emission.
Since we intend to subtract the pure cluster emission to see if the 
residuals are correlated with the radio structure, the above procedure avoids 
subtracting too much of the extended X-ray emission which originally might be associated 
with the radio source. 

\begin{figure}
\resizebox{\hsize}{!}{\includegraphics{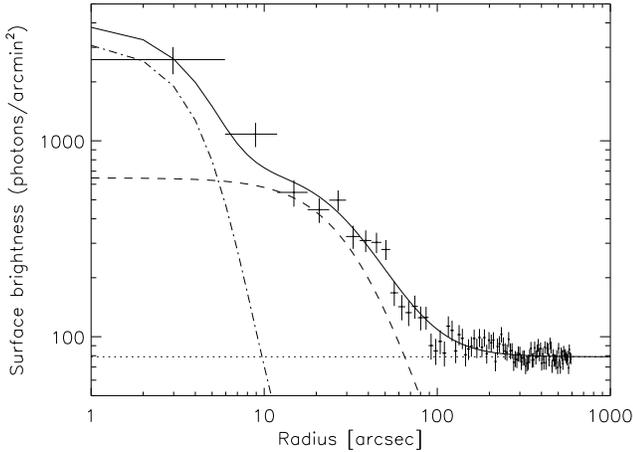}}
\caption{The surface brightness profile for Hercules A. The best-fit $\beta$-model
is indicated by the long-dashed line, whereas the background level and the the PSF
model for the central point source are given by the dotted and the dot-dashed lines
respectively.}
\label{profile}
\end{figure}

The surface brightness profile, shown in Fig.~\ref{profile}, peaks at the position of 
the AGN. In the fitting 
procedure we therefore only used data points for radial distances greater than
10 arcsec to avoid contamination by AGN emission. The best-fit values for the 
$\beta$--model are $S_0 = 647.1_{-100.5}^{+101.0}$ cts/arcmin$^2$, $\beta = 0.63_{-0.03}^{+0.05}$ 
and $r_c = 34.6_{-3.3}^{+3.1}$ arcsec (corresponding to 120.4$_{-11.5}^{+10.8}$ kpc). 
To account for the central AGN emission we included the Point-Spread-Function (PSF) model 
(David et al. 1997) for the HRI in the fit. The theoretical PSF-model was convolved 
with an additional 
Gaussian ($\sigma = 1.5$ arcsec) to take into account the known smearing of the PSF by 
residual wobble motion, which varies between different observations (Morse 1994). 
For the normalisation of the PSF model we get $3973.8_{-1090.0}^{+2790.0}$ cts 
arcmin$^{-2}$.
By integrating the two profiles we calculate the 
contribution of the AGN to the total X-ray emission to be about 8 per cent.
Assuming a power law spectrum with $\Gamma = 1.7$ and Galactic absorption we get a
0.1--2.4 keV flux from the point source of $f_{\rm x} \approx 3.3\times 10^{-13}$ 
erg s$^{-1}$ cm$^{-2}$ which corresponds to a rest frame 0.1--2.4 luminosity of 
$3.4\times 10^{43}$ erg s$^{-1}$. \hera has been noted to be over-luminous in X-rays
with respect to its optical continuum and line luminosities (Siebert et al. 1996;
Brinkmann et al. 1995). However, these investigations were based on the ROSAT All-Sky
Survey data for \hera and no separation of extended and AGN emission could be done.
The much lower AGN luminosity obtained from our spatial analysis places \hera with
the bulk of radio galaxies investigated in the above mentioned studies.
The flux and luminosity for the extended emission 
component are $f_{\rm x} \approx 4.1\times 10^{-12}$ erg s$^{-1}$ cm$^{-2}$ and
$L_{\rm x} \approx 4.3\times 10^{44}$ erg s$^{-1}$ (assuming a thermal Bremsstrahlung
spectrum of $kT = 4.3$ keV; see below). The extended X-ray luminosity
is typical for clusters of galaxies with Abell richness class two or higher 
(Ebeling 1993). This would also be consistent with the number counts of Allington-Smith
et al. (1993) to determine the galaxy density around Hercules A. In addition, the 
observed luminosity and temperature for \hera are consistent with the well known 
luminosity-temperature relation for clusters of galaxies (David et al. 1993).
 
To investigate the residual X-ray emission from the AGN we subtracted the best-fit
one dimensional $\beta$-model given above from the binned and smoothed HRI image. 
The resulting X-ray contours overlayed on an optical image from the corresponding 
POSS-I plate are shown in Fig.~\ref{opt}

\begin{figure}
\vskip8cm
\caption{Residual X-ray contours after subtracting the best-fit $\beta$-model
from the HRI image overlayed on an optical image of the field from the digitized POSS-I
plates. The contours correspond to 3,5,8,12,16,24 and 32 $\sigma$ above background. 
The lowest and highest contour denote intensity levels of $7.0\times 10^{-3}$ 
and $40.3\times 10^{-3}$ cts s$^{-1}$ arcmin$^{-2}$ respectively.} 
\label{opt}
\end{figure}

The residual X-ray emission is dominated by the point source which is most likely
associated with the active nucleus of Hercules A. Several other sources contribute
to the remaining X-ray emission. A significant fraction of the diffuse component
might still be unsubtracted cluster emission, since we assumed a circular symmetry
in the cluster subtraction procedure. Part of the emission might also
come from additional point sources, for example the bright ($m_{\rm V} = 11.97$) 
stellar object to the northwest of Hercules A. Other features, as for example the one-sided 
jet-like emission and the diffuse component about 1.5 arcmin to the west of Hercules A,
correlate with the radio emission. They will be discussed in detail in Sect.~3.4.  
 
\subsection{Physical parameters of the Cluster emission}

By deprojection of the surface brightness profile one can derive the corresponding 
density profile:
\begin{equation}
n(r) = n_0 \left(1+\frac{r^2}{r_c^2}\right)^{-\frac{3\beta}{2}}
\end{equation}
The central density $n_0$ is then given by (e.g. Henry et al. 1993):
\begin{eqnarray*}
n_0 & = & 1.2\times 10^{12} \mbox{cm}^{-3} \left[\frac{I_0}{r_c(kT)^{1/2}}\right]^{1/2} \times \\
    &   & \times \left\{\frac{\Gamma(3\beta-1/2)}{\Gamma(3\beta)}
          \left[\gamma\left(0.7,\frac{E_2}{kT}\right) 
                - \gamma\left(0.7,\frac{E_1}{kT}\right)\right]\right\}^{-1/2}
\end{eqnarray*}
$I_{\rm 0}$ is the central intensity $S_0$ converted to erg cm$^{-2}$ s$^{-1}$ sr$^{-1}$, 
$r_{\rm c}$ the core radius in cm and $kT$ is given in keV. $\Gamma$ and $\gamma$ are the 
complete and incomplete Gamma functions. $E_1$ and $E_2$ denote the observed energy range, 
i.e. 0.1 and 2.4 keV in our case. For the conversion from $S_0$ to $I_0$ we assumed a thermal
bremsstrahlung spectrum with $kT = 4.3$ keV and Galactic absorption. Using the best-fit
parameters from our $\beta$-model we obtain $n_0 = 9.1\times 10^{-3}$ cm$^{-3}$. 
By integrating Eq.~(1), we can now calculate the total gas mass within a given radius.
In the case of \hera we get  $1.6 \times 10^{13}$ M$_{\sun}$ within six times the core radius,
i.e. about 600 kpc. 

Assuming hydrostatic equilibrium for the intracluster medium of Hercules A and spherical 
symmetry, we derive the total gravitating mass as a function of radius from the temperature
and gas density profiles by applying the hydrostatic equation:
\begin{equation}
M_{\rm tot}(r) = -\frac{k T_{\rm g}(r) r}{m_{\rm H} \mu G}
\left(\frac{r}{T_{\rm g}(r)}\frac{dT_{\rm g}(r)}{dr} + \frac{r}{\rho(r)}\frac{d\rho(r)}{dr}\right)
\end{equation}
As discussed in B\"ohringer et al. (1998), deviations from hydrostatic equilibrium as well 
as moderate ellipticities do not have a large effect on the mass determination. With the
additional assumption of an isothermal intracluster medium, Eq.~(2) reduces to
\begin{equation}
M_{\rm tot}(r) = \frac{3\beta k T_{\rm g}}{G\mu m_{\rm H} r_{\rm c}^2} 
                \frac{r^3}{1+\frac{r^2}{r_{\rm c}^2}} 
\end{equation}
For the total gravitating mass within 600 kpc we get $8.4\times 10^{13}$ M$_{\sun}$. Hence, 
the gas mass fraction is $\approx 18$ per cent at this radius. 

\subsection{Radio/X-ray interaction}

Fig.~\ref{radiox} shows the radio contours from a 5GHz VLA observation (Dreher \& Feigelson 
1984), overlayed onto the HRI image after subtraction of the cluster emission according to 
the best-fit $\beta$-model. Several features of the residual X-ray emission
coincide spatially with the radio emission: (a) faint knots of X-ray emission are 
surrounding the outer edges of both radio lobes; (b) there is a bright patch of X-ray 
emission close to 
the head of the western radio lobe. It is only slightly displaced from a faint radio
feature, which might be interpreted as the head of the advancing radio jet; (c) the 
orientation of the jet-like feature in the X-ray emission close to the central AGN is 
aligned with the radio jet; (d) the brightest spot in the eastern radio jet {\bf is bracketed
by} enhancements in the X-ray emission. We note that all features are robust
in the sense that they also show up when we subtract the cluster emission determined
from the azimuthally averaged radial profile.
  
\begin{figure*}
\vskip9cm
\caption{Radio contours overlayed onto the HRI image after subtraction of the
best-fit $\beta$-model. The radio map was obtained with the VLA at 5GHz (Dreher \&
Feigelson 1984).}
\label{radiox}
\end{figure*}

The close association of X-ray and radio features strongly argues for interaction
of the relativistic gas in the radio jet and the thermal gas of the intracluster 
medium. An X-ray cavity coinciding with the radio lobes has been noted previously 
for the radio galaxies NGC 1275 (B\"ohringer et al. 1993) and Cygnus A (Carilli et al. 
1994). In the latter case also an enhancement of the X-ray emission close to the
edges of the radio lobes was reported. The general physical scenario for these effects
(cf. Carilli et al. 1994 {\bf and Clarke et al. 1997 for a more detailed treatment}) is such 
that the advancing and expanding jet is ploughing 
its way through the intracluster medium (ICM) thereby expelling the thermal gas from 
its interior. A thin sheath of dense, shocked material develops in the vicinity of 
the advancing head and the expanding tails of the radio lobes. When our line of sight 
is tangential to the edges of the radio lobes, these density enhancements become 
visible in X-rays, because thermal X-ray emission is proportional to the square of 
the gas density. For lines of sight through the center of the radio lobes the
X-ray enhancements are probably balanced by an X-ray cavity within the radio lobes
due to the expelled gas. The faint patches of X-ray emission surrounding the radio 
lobes in \hera and the bright spot close to the head of the western lobe qualitatively
fit into this scenario. 

The origin of the jet-like feature in the residual X-ray image is unclear. One might
speculate that it is due to synchrotron-self Compton emission from the radio jet. The
strength of this 'X-ray jet' and the absence of any counter jet would imply that
\hera is not oriented in the plane of the sky. This obviously is in contradiction to
the existence of a double radio jet and the overall symmetry of the radio source (Dreher
\& Feigelson 1984). 

Another interesting feature to note is the close association of the brightest knot 
in the eastern radio jet and the strongly enhanced X-ray emission to the south (and, 
at a much lower brightness level, to the north). It almost seems as if the radio jet
is squeezed in and probably deflected by local enhancements of the surrounding gas 
density. But again an unambiguous interpretation is difficult in our case, because the 
maximum of the X-ray emission is close to a rather bright stellar object on the optical 
plate, presumably a foreground star, which might contribute significantly to the X-ray 
flux.    

Since X-rays from radio lobes due to inverse Compton scattered cosmic microwave 
background (CMB) photons are mandatory, we also estimated the amount of X-ray emission 
expected from this effect. Applying the formalism described in Feigelson et al. (1995), 
we first determined the magnetic field within the radio lobes from equipartition 
arguments. As a conservative lower limit we get $\approx 10 \mu$G. Using this value 
for the magnetic field we estimate that at most a flux of F$_{\rm 0.1-2.4 keV}\approx 
4\times 10^{-15}$ erg cm$^{-2}$ s$^{-1}$ is expected from this process. Since this 
flux level is about a factor of 20 below the background flux level in our observation, 
there is no chance to detect these X-rays.

\section{Summary and conclusions}

We investigated the spectral and spatial properties of the X-ray emission of the
prominent radio galaxy \hera using ROSAT HRI, PSPC, and ASCA observations. The 
ASCA data clearly favor a thermal plasma model with a temperature 
of about 4.3 keV and abundances $\approx 0.4$ solar. A small ($\approx 8$ per 
cent) non-thermal contribution of the central AGN is consistent with the ASCA data.

The X-ray source associated with \hera is clearly extended in the ROSAT HRI 
observation. In addition, it is slightly elongated in the direction of the radio 
jets and lobes. The outer part of the surface brightness profile perpendicular to 
the radio structure is well described by a King--model with $\beta = 0.63$ and 
a core radius of about 120 kpc. The total mass within 600 kpc is $8.4\times 10^{13}$
M$_{\sun}$ and the gas mass fraction is about 18 per cent. A central 
point source contributing about 8 per cent to the total X-ray flux is clearly 
needed to fit the surface brightness distribution. We determine a luminosity of 
$L_{\rm 0.1-2.4 keV} = 3.4\times 10^{43}$ erg s$^{-1}$ for the AGN and $L_{\rm 0.1-2.4 keV} 
= 4.3\times 10^{44}$ erg s$^{-1}$ for the extended emission.

After subracting the extended X-ray emission, significant residuals remain in 
the X-ray image, which partly coincide with the radio emission. They strongly
suggest an interaction of the radio jet with the intracluster gas, an effect which
has already been noted previously for Cygnus A (Carilli et al. 1994), NGC 1275 
(B\"ohringer et al. 1993) and MRC 0625-536 (Otani et al. 1998).     
   
Spectral and spatial analysis thus provides clear evidence that \hera is indeed 
located in a cluster of galaxies, but spectroscopic observations are needed to 
unambigously confirm this interpretaion. The non-spherical shape of the extended 
X-ray emission indicates that the cluster is not relaxed yet, a view which is 
supported by the identification of a double nucleus in the central cD galaxy, thus
indicating an ongoing merging process.   
  
\begin{acknowledgements}
JS acknowledges financial support from the RIKEN--MPG exchange program. 
JS also thanks his colleagues from the Cosmic Radiation Laboratory at RIKEN 
for hospitality and support during a stay at the institute, where part of 
this work was done. It is a pleasure to thank Eric Feigelson for providing 
the radio contour map. This research has made use of the NASA/IPAC Extragalactic 
Data Base (NED), which is operated by the Jet Propulsion Laboratory, 
California Institute of Technology, under contract with the National 
Aeronautics and Space Administration.
\end{acknowledgements}

\end{document}